\documentclass[aps,prl,reprint,superscriptaddress,showpacs,longbibliography]{revtex4-1}
\usepackage{mathrsfs}
\usepackage{amsmath}
\usepackage{amsfonts}
\usepackage{amssymb}
\usepackage{amsthm}
\usepackage{graphicx}
\usepackage{natbib}
\usepackage{color}
\usepackage{hyperref}
\usepackage{bm}
\usepackage[caption=false]{subfig}
\usepackage{verbatim}

\begin{document}
\title{Noninteger-spin magnonic excitations in untextured magnets}
\author{Akashdeep Kamra}
\email{akashdeep.kamra@uni-konstanz.de}
\affiliation{Department of Physics, University of Konstanz, D-78457 Konstanz, Germany}
\author{Utkarsh Agrawal}
\affiliation{Department of Physics, Indian Institute of Technology Bombay, Powai, Mumbai-400076, India}
\author{Wolfgang Belzig}
\email{wolfgang.belzig@uni-konstanz.de}
\affiliation{Department of Physics, University of Konstanz, D-78457 Konstanz, Germany}

\begin{abstract}
 Interactions are responsible for intriguing physics, e.g. emergence of exotic ground states and excitations, in a wide range of systems. Here we theoretically demonstrate that dipole-dipole interaction leads to bosonic eigen-excitations with average spin ranging from zero to above $\hbar$ in magnets with uniformly magnetized ground states. These exotic excitations can be interpreted as quantum coherent conglomerates of spin $\hbar$ magnons, the eigen-excitations when the dipolar interactions are disregarded. We further find that the eigenmodes in an easy-axis antiferromagnet are spin-zero quasiparticles instead of the widely believed spin $\pm \hbar$ magnons. The latter re-emerge when the symmetry is broken by a sufficiently large applied magnetic field. The average spin greater than $\hbar$ is accompanied by vacuum fluctuations and may be considered to be a weak form of frustration. 
\end{abstract} 




\maketitle


{\it Introduction.} Interactions among the entities constituting a physical system determine its qualitative behavior. Even a simple metal, typically described as a free (non-interacting) electron gas, comprises an electronic Fermi liquid~\cite{Brown1971} with the quasiparticle effective mass subtly absorbing the interaction effects. These quasiparticles however maintain same charge ($e$) and spin ($\hbar / 2$) as that of a single electron. Relatively exotic states are exemplified by the superconducting phase, in which phonon-mediated electron-electron interaction leads to the formation of Cooper pairs with charge 2$e$, and the fractional quantum hall phase~\cite{Laughlin1983} composed of fractional-charge quasiparticles~\cite{Jain1989,Saminadayar1997,Reznikov1999}. 

In a ferromagnet (FM), the exchange interaction favors a spin-aligned ground state. Considering only Zeeman and exchange interactions, the eigen-excitations are charge-neutral, spin $\hbar$ bosonic quasiparticles - magnons~\cite{Kittel1963}, where the role of exchange is again absorbed in the magnon effective mass. Magnetic dipolar interaction (DI)~\cite{dipnote} between these otherwise ``non-interacting'' magnons lead to formation of new quasiparticles - squeezed magnons~\cite{Kamra2016A}. Furthermore, dipolar and spin-orbit interactions often result in textured ground states, such as domain walls~\cite{Kittel1949} and skyrmions~\cite{Bogdanov2001,Wiesendanger2016}, exhibiting rich physics.

Analogously, antiferromagnets~\footnote{Strictly speaking, we consider easy-axis AFs in which the spins prefer to align along a certain direction.} (AFs) admit the so-called Neel-ordered ground state with two interpenetrating, equivalent, and anti-collinear sublattices resulting in a vanishing net magnetization. The magnetic order without any associated magnetization and stray fields has led to a strong interest in employing AFs for spintronic applications~\cite{Gomonay2014,Jungwirth2016} with demonstrated success~\cite{Wadley2016}. The normal excitations in AFs are believed to be similar in nature to the magnons in FMs~\cite{Kittel1963}, with the former kind's much larger energies being dominated by the inter-sublattice antiferromagnetic exchange. Typically, this large energy argument is employed to justify disregarding DI in considering AF magnons~\cite{Keffer1952,Ohnuma2013,Gomonay2014,Cheng2014,Rezende2016}. While there is some merit to the energy comparison, insofar as accounting for DI leaves the AF magnon energies essentially unchanged, the concomitant breaking of the rotational invariance of the spin system leads to qualitatively important and novel effects that form a part of the results reported herein.

Apart from the ordered configurations discussed above, magnets also allow phases, called spin liquids~\cite{Balents2010,Savary2016}, with long-range spin correlations without any static spin-order. This intriguing phenomenon, called geometrical frustration, may occur when each spin faces conflicting demands from its neighbors as to the direction in which the spin should point, for example in a triangular lattice AF. A static spin-order is absent in such a system because it fluctuates between a multitude of degenerate states with opposite net spins. Spin-liquids are believed to host several exotic excitations~\cite{Balents2010,Savary2016} such as magnetic monopoles~\cite{Castelnovo2008}, spinons~\cite{Affleck1988,Read1991,Rantner2001}, triplons etc. and are speculated as a possible precursor to high-$\mathrm{T}_\mathrm{C}$ superconductivity~\cite{Anderson1987,Lee2006}. While nearest-neighbor exchange models are often employed to capture the essential physics, the DI plays an important role in some spin-liquids, for instance in spin-ice systems~\cite{Harris1997,Bramwell2001}, in which magnetic monopoles have recently been observed~\cite{Morris2009,Kadowaki2009}.

\begin{figure*}[th]
\begin{center}
\subfloat[Quasiferromagnet $M_{\mathcal{A}0} = 5 M_{\mathcal{B}0}$]{\includegraphics[width=58.5mm]{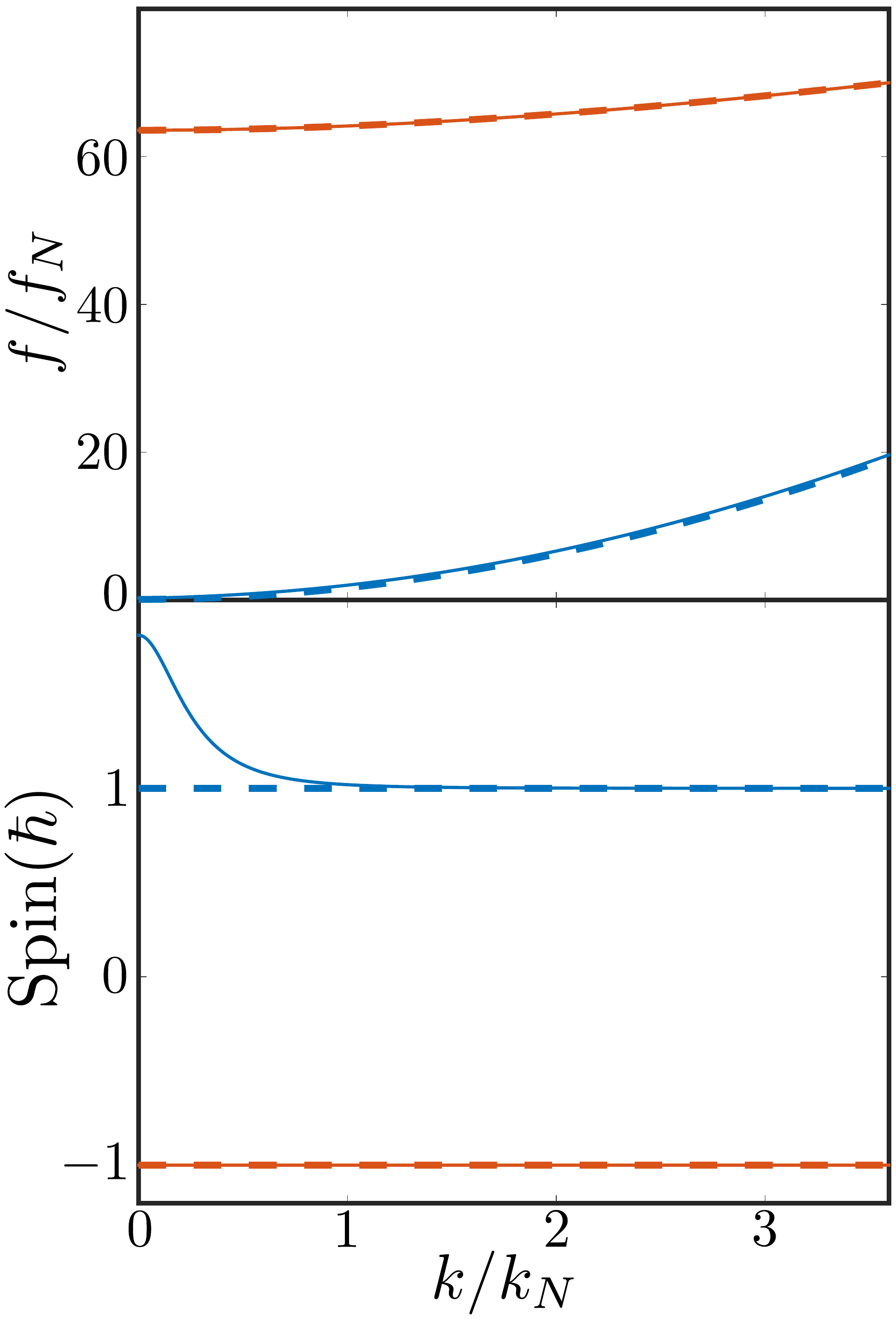}} \
\subfloat[Ferrimagnet $M_{\mathcal{A}0} = 2 M_{\mathcal{B}0}$]{\includegraphics[width=55mm]{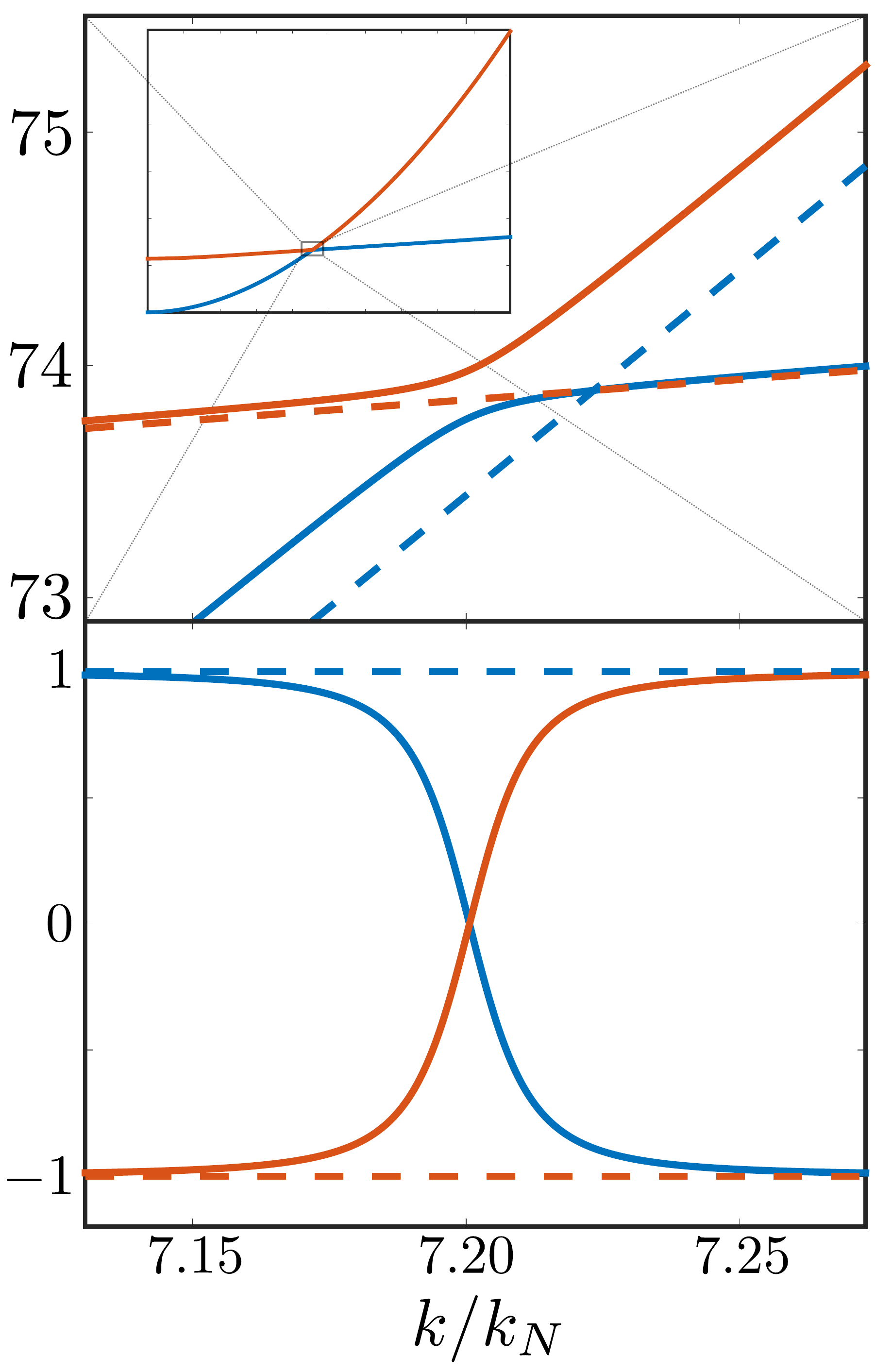}} \
\subfloat[Antiferromagnet $M_{\mathcal{A}0} = M_{\mathcal{B}0}$]{\includegraphics[width=55mm]{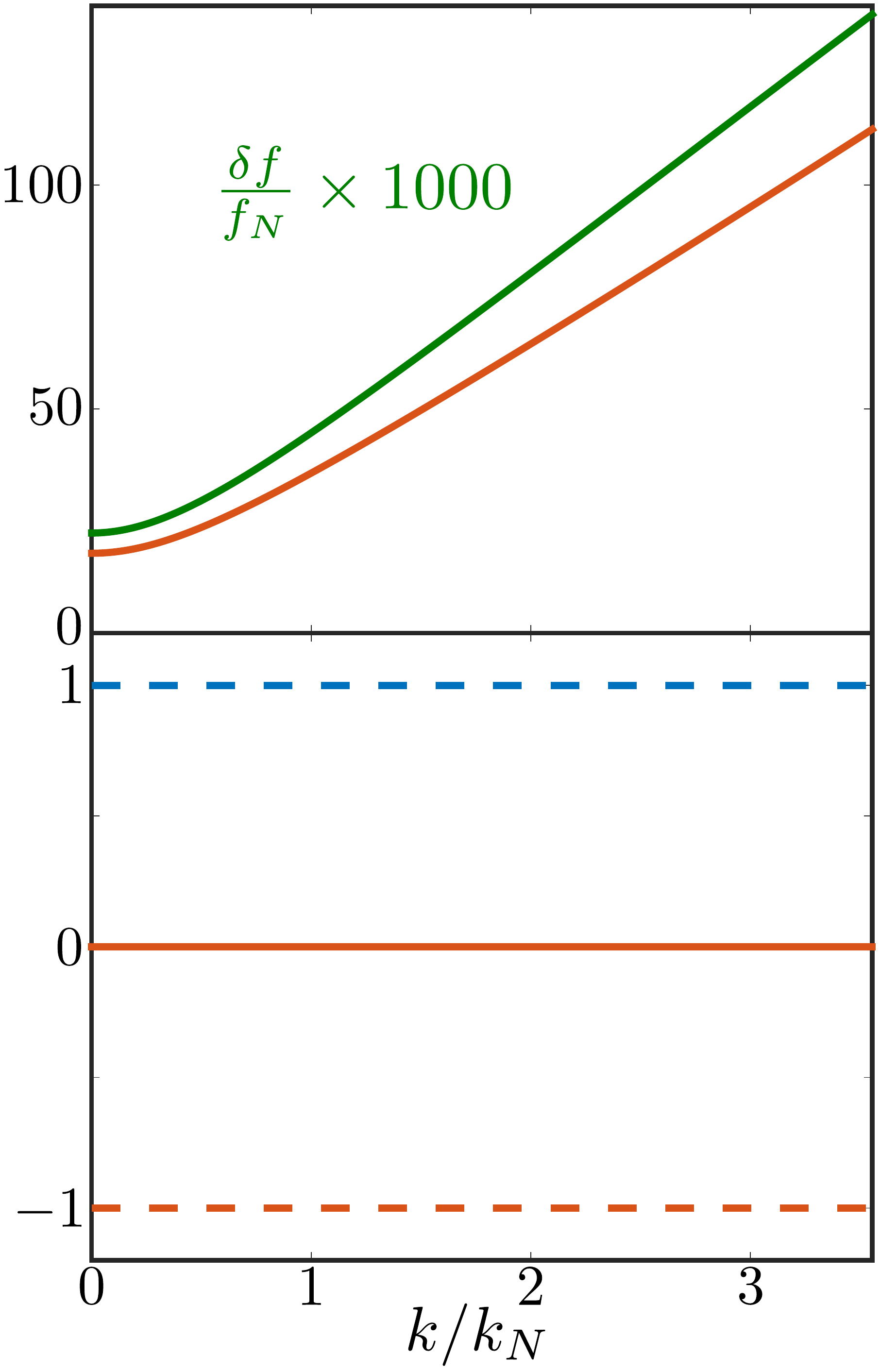}}
\caption{Dispersion along x-direction (upper panels) and quasiparticle spin (lower panels) for three kinds of magnets. Solid and dashed lines respectively denote the results with and without DI. $2 \pi f_N = |\gamma_{\mathcal{A}}| \mu_0 M_{\mathcal{A}0}$ and $f_{l}(k_N) = 2 \max[f_N,f_{l}(0)]$ define the normalizations $f_N,~k_N$ with $f_{l}(k)$ the lower dispersion band. (a) The low energy band mimics magnon dispersion in a ferromagnet. Quasiparticles at low (comparable to dipolar) energies exhibit squeezing and spin greater than $\hbar$~\cite{Kamra2016A}. (b) Region around the dispersion anticrossing point is shown with the upper panel inset depicting the full dispersion. Quasiparticles with spin varying between $+\hbar$ and $-\hbar$ are formed around the anticrossing point. (c) Degeneracy of the two bands in antiferromagnets is lifted by DI with splitting $\delta f$ in the GHz range. Quasiparticle spin in both bands vanishes.}
\label{fig:ferro}
\end{center}
\end{figure*}

In this Rapid Communication, we theoretically investigate excitations in magnets with homogeneous spin-ordered ground states laying special emphasis on the role of DI. The qualitative physics presented herein relies upon the presence of a spin-non-conserving interaction, e.g. DI. While a general framework for DI is presented here, depending upon the material, the spin-non-conserving contribution may predominantly be of a different physical origin such as magnetocrystalline anisotropy. The easy-axis two-sublattice model employed herein encompasses the full range from FMs via ferrimagnets to AFs. Our key finding is that DI relaxes the spin conservation constraint and results in exotic bosonic quasiparticles with non-integral average spin~\footnote{In the following, we use the terms ``spin'' and ``average spin'' interchangeably.} ranging from zero to greater than $\hbar$ (Fig. \ref{fig:ferro}). These new quasiparticles can be physically understood as quantum coherent conglomerates of spin $\pm \hbar$ magnons. The quasiparticles with spin greater and smaller than $\hbar$ are qualitatively different in nature, and result, respectively, from squeezing and hybridization dominated coherences. While spin less than $\hbar$ excitations require hybridization between spin $+\hbar$ and $-\hbar$ magnons, squeezed excitations can qualitatively be understood within a single sublattice model, and are constituted by a  coherent superposition of different number states of the spin $\hbar$ magnon.

\begin{figure}[t]
\begin{center}
\includegraphics[width=70mm]{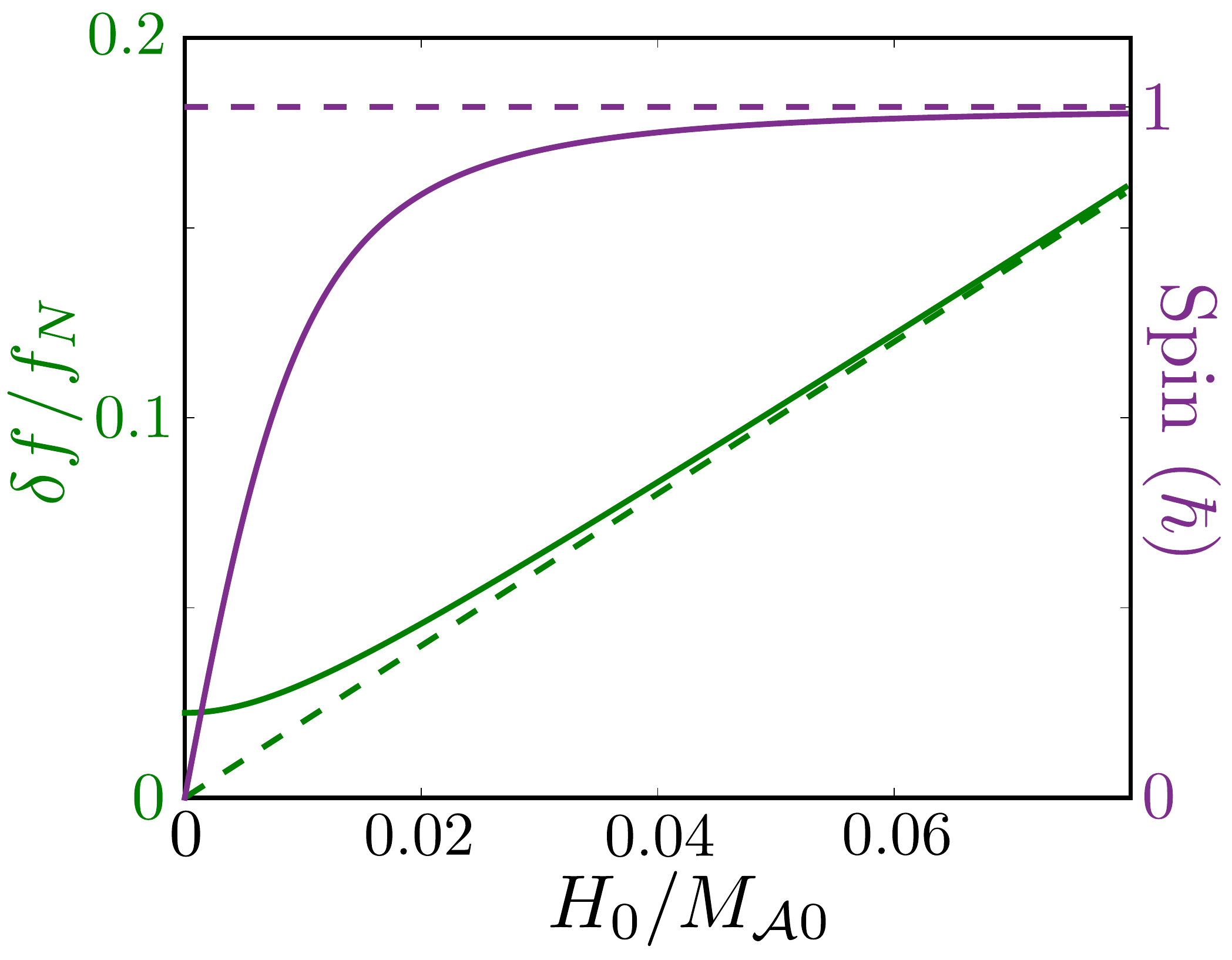}
\caption{Normalized band spitting ($2 \pi f_N = |\gamma_{\mathcal{A}}| \mu_0 M_{\mathcal{A}0}$) and quasiparticle spin vs. applied magnetic field ($H_0$) in easy-axis antiferromagnets. Solid and dashed lines respectively denote the results with and without DI.}
\label{fig:split}
\end{center}
\end{figure}

Specifically, we find that quasiparticles in AFs with equivalent sublattices possess zero average spin and are non-degenerate [Fig. \ref{fig:ferro} (c)], in contrast with the wide belief based upon theories that disregard DI~\cite{Keffer1952,Ohnuma2013,Gomonay2014,Cheng2014,Rezende2016}. These quasiparticles continuously approach spin $\pm \hbar$ magnons, due to broken sublattice equivalence, as the applied magnetic field becomes much larger than the sublattice saturation magnetization (Fig. \ref{fig:split}). The squeezing-mediated quasiparticle spin greater than $\hbar$ is inevitably accompanied by vacuum fluctuations which decrease the net spin in the ground state. We argue that this behavior is a consequence of large ground state degeneracy, and is thus a weak, non-geometrical form of frustration that only diminishes the ground state net spin. The spin greater than $\hbar$ of squeezed magnons has already been discussed in the context of spin transport~\cite{Kamra2016A}. Here, we highlight a different property, i.e. relation to frustration, of these quantum excitations.


{\it Theoretical model.} We start by outlining the method. The first step is formulating the classical free energy density of two antiferromagnetically coupled sublattices $\mathcal{A}$ and $\mathcal{B}$ in terms of their respective magnetizations $\pmb{M}_{\mathcal{A,B}}(\pmb{r})$~\cite{Akhiezer1968}. With ferromagnetic intrasublattice exchange and easy-axis anisotropy along z-direction, the ground state is Neel ordered, i.e. $\pmb{M}_{\mathcal{A,B}}(\pmb{r}) = \pm M_{\mathcal{A}0,\mathcal{B}0} \hat{\pmb{z}}  + \pmb{\delta M}_{\mathcal{A,B}}(\pmb{r})$. Here $M_{\mathcal{A}0,\mathcal{B}0} $ are the respective sublattice saturation magnetizations, and $\pmb{\delta M}_{\mathcal{A,B}}(\pmb{r})$ describe the excitations with $|\pmb{\delta M}_{\mathcal{A,B}}(\pmb{r})| \ll M_{\mathcal{A}0,\mathcal{B}0}$. The quantum Hamiltonian density is then obtained by replacing classical [$\pmb{M}_{\mathcal{A,B}}(\pmb{r})$] with quantum~\footnote{Here, and in the rest of the manuscript, we employ tilde to represent quantum operators.} field variables $\tilde{\pmb{M}}_{\mathcal{A,B}}(\pmb{r})$~\cite{Kittel1963,Akhiezer1968}. The Holstein-Primakoff transformations~\cite{Holstein1940,Kittel1963,Akhiezer1968} substitute $\tilde{\pmb{M}}_{\mathcal{A,B}}(\pmb{r})$ in terms of the bosonic ladder operators $\tilde{a}(\pmb{r}), \tilde{b}(\pmb{r}^\prime)$ that, respectively, represent annihilation of a magnon on the sublattices $\mathcal{A},\mathcal{B}$ at positions $\pmb{r},\pmb{r}^\prime$. Integrating the ensuing Hamiltonian density over the entire volume yields the system Hamiltonian in terms of the Fourier space magnon operators $\tilde{a}_{\pmb{k}}, \tilde{b}_{\pmb{k}^\prime}$. Finally, the eigen-excitations and the corresponding dispersion are obtained by diagonalizing the Hamiltonian using a four-dimensional Bogoliubov transform.

The classical free energy density for the two-sublattice system described above comprises of Zeeman ($H_{\mathrm{Z}}$), anisotropy ($H_{\mathrm{an}}$), exchange ($H_{\mathrm{ex}}$), and dipolar ($H_{\mathrm{dip}}$) interaction terms. With an applied magnetic field $H_0 \hat{\pmb{z}}$ and $\mu_0$ the permeability of free space, the Zeeman energy density reads $H_{\mathrm{Z}} = - \mu_0 H_0 (M_{\mathcal{A}\mathrm{z}} + M_{\mathcal{B}\mathrm{z}})$. The easy-axis anisotropy is parametrized in terms of the constants $K_{\mathrm{u}\mathcal{A}},~K_{\mathrm{u}\mathcal{B}}$ as $H_{\mathrm{an}} = - K_{\mathrm{u}\mathcal{A}} M_{\mathcal{A}\mathrm{z}}^2 - K_{\mathrm{u}\mathcal{B}} M_{\mathcal{B}\mathrm{z}}^2$~\cite{Akhiezer1968}. In a minimalistic model, the exchange energy density is expressed in terms of the constants $\mathcal{J}_{\mathcal{A}},~\mathcal{J}_{\mathcal{B}},~\mathcal{J}_{\mathcal{AB}}$ and $\mathcal{J}$ as follows~\cite{Akhiezer1968}: $H_{\mathrm{ex}} = \sum_{x_i = x,y,z} [ \mathcal{J}_{\mathcal{A}}  (\partial \pmb{M}_{\mathcal{A}}/ \partial x_{i}) \cdot (\partial \pmb{M}_{\mathcal{A}}/ \partial x_{i}) + \mathcal{J}_{\mathcal{B}} (\partial \pmb{M}_{\mathcal{B}}/ \partial x_{i}) \cdot (\partial \pmb{M}_{\mathcal{B}}/ \partial x_{i}) + \mathcal{J}_{\mathcal{AB}} (\partial \pmb{M}_{\mathcal{A}}/ \partial x_{i}) \cdot (\partial \pmb{M}_{\mathcal{B}}/ \partial x_{i})] + \mathcal{J} \pmb{M}_{\mathcal{A}} \cdot \pmb{M}_{\mathcal{B}}$. The DI energy density is obtained in terms of the demagnetization field $\pmb{H}_{m}$~\cite{Akhiezer1968,Kittel1963}: $H_{\mathrm{dip}} = - (1/2) \mu_0 \pmb{H}_{m} \cdot (\pmb{M}_{\mathcal{A}} + \pmb{M}_{\mathcal{B}})$. The demagnetization field is related to the magnetization via the Maxwell's equations simplified within the magnetostatic approximation~\cite{Akhiezer1968,Kittel1963,SupplMat}.  

In order to quantize the Hamiltonian, we recognize the relation between magnetization and spin density operators: $\tilde{\pmb{M}}_{\mathcal{A,B}}(\pmb{r}) = - |\gamma_{\mathcal{A,B}}| \tilde{\pmb{S}}_{\mathcal{A,B}}(\pmb{r})$, where $\gamma_{\mathcal{A,B}}$ are the respective gyromagnetic ratios assumed to be negative here. The commutation relations for the magnetization components can be obtained using spin commutation rules, and are satisfied by the Holstein-Primakoff transformations~\cite{Holstein1940} generalized to the two sub-lattice model~\cite{Kittel1963,Ohnuma2013}. To the lowest order, the latter are given by: $\tilde{M}_{\mathcal{A}+}(\pmb{r}) = \sqrt{2 |\gamma_{\mathcal{A}}| \hbar M_{\mathcal{A}0}} ~ \tilde{a}(\pmb{r})$, $\tilde{M}_{\mathcal{B}+}(\pmb{r}) = \sqrt{2 |\gamma_{\mathcal{B}}| \hbar M_{\mathcal{B}0}} ~ \tilde{b}^{\dagger}(\pmb{r})$, $\tilde{M}_{\mathcal{A}\mathrm{z}}(\pmb{r}) = M_{\mathcal{A}0} - \hbar |\gamma_{\mathcal{A}}| \tilde{a}^{\dagger}(\pmb{r}) \tilde{a}(\pmb{r})$, and $\tilde{M}_{\mathcal{B}\mathrm{z}}(\pmb{r}) = - M_{\mathcal{B}0} + \hbar |\gamma_{\mathcal{B}}| \tilde{b}^{\dagger}(\pmb{r}) \tilde{b}(\pmb{r})$. Here, $\tilde{M}_{\mathcal{A}+} = \tilde{M}_{\mathcal{A}-}^{\dagger} = \tilde{M}_{\mathcal{A}\mathrm{x}} + i (\gamma_{\mathcal{A}}/|\gamma_{\mathcal{A}}|) \tilde{M}_{\mathcal{A}\mathrm{y}}$, and similar for the sublattice $\mathcal{B}$. The final step in obtaining the Hamiltonian is integrating the energy density operator over the volume. 

Delegating details to the supplemental material~\cite{SupplMat}, the Hamiltoninan is thus obtained in terms of the k-space magnon ladder operators $\tilde{a}_{\pmb{k}}, \tilde{b}_{\pmb{k}}$:
\begin{align}\label{eq:Hamil1}
\tilde{\mathcal{H}}  = & \sum_{\pmb{k}} \left[ \frac{A_{\pmb{k}}}{2} \tilde{a}_{\pmb{k}}^{\dagger} \tilde{a}_{\pmb{k}} + \frac{B_{\pmb{k}}}{2} \tilde{b}_{\pmb{k}}^{\dagger} \tilde{b}_{\pmb{k}} + C_{\pmb{k}} \tilde{a}_{\pmb{k}} \tilde{b}_{-\pmb{k}} + D_{\pmb{k}} \tilde{a}_{\pmb{k}} \tilde{a}_{-\pmb{k}} \right. \nonumber \\
      &  \left.+ E_{\pmb{k}} \tilde{b}_{\pmb{k}} \tilde{b}_{-\pmb{k}} + F_{\pmb{k}} \tilde{a}_{\pmb{k}} \tilde{b}_{\pmb{k}}^{\dagger} \right] + \mathrm{h.c.} \quad .
\end{align}
Analytical expressions for the coefficients $A_{\pmb{k}}, B_{\pmb{k}} \cdots$ are given in the supplemental material~\cite{SupplMat}. Here, we simply note that $C_{\pmb{k}}$ is dominated by intersublattice exchange, while $D_{\pmb{k}}, E_{\pmb{k}}, F_{\pmb{k}}$ result entirely from DI. All the coefficients stay unchanged on replacing $\pmb{k}$ with $-\pmb{k}$. 

In order to diagonalize the Hamiltonian, it is convenient to define a new wavevector index $\pmb{\kappa}$ that runs over only half of the wavevector space~\cite{Holstein1940}. We define a four-dimensional Bogoliubov transform to new bosonic operators $\tilde{\alpha}_{\pm\pmb{\kappa}}, \tilde{\beta}_{\pm\pmb{\kappa}}$:
\begin{align}\label{eq:S1}
\begin{pmatrix}
\tilde{\alpha}_{\pmb{\kappa}} \\ 
\tilde{\beta}_{-\pmb{\kappa}}^{\dagger} \\
\tilde{\alpha}_{-\pmb{\kappa}}^{\dagger} \\
\tilde{\beta}_{\pmb{\kappa}}
\end{pmatrix}
 = & 
 \begin{pmatrix}
 u_{1  } & v_{1  } & w_{1  } & x_{1  } \\
 u_{2  } & v_{2  } & w_{2  } & x_{2  } \\
 u_3 & v_3 & w_3 & x_3 \\
 u_4 & v_4 & w_4 & x_4
 \end{pmatrix}
 \begin{pmatrix}
  \tilde{a}_{\pmb{\kappa}} \\ 
  \tilde{b}_{-\pmb{\kappa}}^{\dagger} \\
  \tilde{a}_{-\pmb{\kappa}}^{\dagger} \\
  \tilde{b}_{\pmb{\kappa}}
  \end{pmatrix}  =
  \underline{S} 
  \begin{pmatrix}
 \tilde{a}_{\pmb{\kappa}} \\ 
 \tilde{b}_{-\pmb{\kappa}}^{\dagger} \\
 \tilde{a}_{-\pmb{\kappa}}^{\dagger} \\
 \tilde{b}_{\pmb{\kappa}}
 \end{pmatrix},
\end{align} 
with the requirement that it diagonalizes the Hamiltonian to $\tilde{\mathcal{H}}  =   \sum_{\pmb{\kappa}} [\epsilon_{\alpha \pmb{\kappa}} ( \tilde{\alpha}_{\pmb{\kappa}}^{\dagger} \tilde{\alpha}_{\pmb{\kappa}}  + \tilde{\alpha}_{\pmb{-\kappa}}^{\dagger} \tilde{\alpha}_{\pmb{-\kappa}}) +  \epsilon_{\beta \pmb{\kappa}} (\tilde{\beta}_{\pmb{\kappa}}^{\dagger} \tilde{\beta}_{\pmb{\kappa}} + \tilde{\beta}_{\pmb{-\kappa}}^{\dagger} \tilde{\beta}_{\pmb{-\kappa}})]$. In the transformation above, the coefficients (such as $w_1, x_2$ etc.) which combine a creation with an annihilation operator are associated with squeezing while the ones (e.g. $x_1, w_2$ ) that combine creation (or annihilation) operators lead to hybridization. The subscript $\pmb{\kappa}$ of $\underline{S}$ and its elements has been suppressed for brevity. Since $A_{\pmb{\kappa}}, B_{\pmb{\kappa}} \cdots$ are invariant with respect to the replacement $\pmb{\kappa} \to \pmb{- \kappa}$, the same is true for $\epsilon_{\alpha \pmb{\kappa}}$, $\epsilon_{\beta \pmb{\kappa}}$ and $\underline{S}$. This invariance also leads to the relations $S_{22} = S_{11}^*$ and $S_{21} = S_{12}^*$, where $S_{ij}$ are $2 \times 2$ matrix blocks that constitute $\underline{S}$. Bosonic commutation rules for $\tilde{\alpha}_{\pmb{\pm\kappa}}, \tilde{\beta}_{\pmb{\pm\kappa}}, \tilde{a}_{\pmb{\pm\kappa}}, \tilde{b}_{\pmb{\pm\kappa}}$ further impose relations on $\underline{S}$ elements, all of which can be succinctly expressed as:
\begin{align}\label{eq:S3}
\underline{S} ~ \underline{Y} ~ \underline{S}^{\dagger} & = \underline{Y}  \implies  \underline{S}^{-1} = \underline{Y} ~ \underline{S}^{\dagger} ~ \underline{Y}^{-1},
\end{align} 
where $\underline{Y} = \sigma_{z} \otimes \sigma_{z}$, with $\sigma_{z}$ the third Pauli matrix. Employing the diagonal form of $\tilde{\mathcal{H}}$, we obtain $[\tilde{\alpha}_{\pmb{\kappa}}, \tilde{\mathcal{H}}] = \epsilon_{\alpha \pmb{\kappa}} \tilde{\alpha}_{\pmb{\kappa}}$, from which Eqs. (\ref{eq:Hamil1}) and (\ref{eq:S1}) lead to the eigenvalue problem $\underline{T}_{\pmb{\kappa}} \xi_{1 \pmb{\kappa}}  = \epsilon_{\alpha \pmb{\kappa}}  \xi_{1\pmb{\kappa}}$ with
\begin{align}
\underline{T}_{\pmb{\kappa}} = &
\begin{pmatrix}
A_{\pmb{\kappa}} & - C_{\pmb{\kappa}} & -2 D_{\pmb{\kappa}} & F_{\pmb{\kappa}} \\
C_{\pmb{\kappa}} & - B_{\pmb{\kappa}} & - F_{\pmb{\kappa}} & 2 E_{\pmb{\kappa}}^* \\
2 D_{\pmb{\kappa}}^* & - F_{\pmb{\kappa}}^* & -A_{\pmb{\kappa}} & C_{\pmb{\kappa}} \\
F_{\pmb{\kappa}}^* & - 2E_{\pmb{\kappa}} & -C_{\pmb{\kappa}} & B_{\pmb{\kappa}}
\end{pmatrix}, \ 
\xi_{1 \pmb{\kappa}} = 
\begin{pmatrix}
u_1 \\ v_1 \\ w_1 \\ x_1
\end{pmatrix}.
\end{align}
Employing $[\tilde{\beta}_{\pmb{\kappa}}, \tilde{\mathcal{H}}] = \epsilon_{\beta \pmb{\kappa}} \tilde{\beta}_{\pmb{\kappa}}$ in an analogous way yields a similar eigenvalue problem $\underline{T}_{\pmb{\kappa}} \xi_{4 \pmb{\kappa}}  = \epsilon_{\beta \pmb{\kappa}}  \xi_{4 \pmb{\kappa}}$. The multiplicative constants for eigenvectors are fixed by Eq. (\ref{eq:S3}). Since $S_{22} = S_{11}^*$, $S_{21} = S_{12}^*$, knowledge of $\xi_{1 \pmb{\kappa}}, \xi_{4 \pmb{\kappa}}$ provides the complete transformation matrix $\underline{S}$. 

While the eigenvalue problem at hand is analytically solvable, we do not present the unwieldy solution here. Instead, we provide the dispersion and eigenvectors {\it disregarding} DI:
\begin{align}\label{eq:sol1}
\epsilon_{\alpha \pmb{\kappa},\beta \pmb{\kappa}}^{\prime} & = \frac{\pm(A_{\pmb{\kappa}}^{\prime} - B_{\pmb{\kappa}}^{\prime}) + \sqrt{(A_{\pmb{\kappa}}^{\prime} + B_{\pmb{\kappa}}^{\prime})^2 - 4 (C_{\pmb{\kappa}}^{\prime})^2}}{2}, \nonumber \\
\xi_{1 \pmb{\kappa}}^{\prime} & = [q n_4, n_2, 0 , 0]^T; \ \xi_{4 \pmb{\kappa}}^{\prime} = [0, 0, q n_2, n_4]^T,
\end{align}
where $q = C_{\pmb{\kappa}}^{\prime}/|C_{\pmb{\kappa}}^{\prime}|$, $n_{2,4} = (1/\sqrt{2}) \sqrt{1/\sqrt{1 - \delta^2} \mp 1}$, and $\delta = 2C_{\pmb{\kappa}}^{\prime}/(A_{\pmb{\kappa}}^{\prime} + B_{\pmb{\kappa}}^{\prime})$. Here, the primed notation emphasizes that all the coefficients have been evaluated disregarding DI.


{\it Excitation spectrum and properties.} We are primarily interested in the eigen-excitation spin. With the total volume $\mathcal{V}$, z-component of the total spin $\tilde{\pmb{S}}(\pmb{r}) = \tilde{\pmb{S}}_{\mathcal{A}}(\pmb{r}) + \tilde{\pmb{S}}_{\mathcal{B}}(\pmb{r})$ is evaluated using the employed Holstein-Primakoff and Bogoliubov [Eq. (\ref{eq:S1})] transformations:
\begin{align}\label{eq:spin}
\int_{\mathcal{V}} d^3 r \left \langle \tilde{S}_{\mathrm{z}}(\pmb{r}) \right \rangle = &  - \left[ \frac{M_{\mathcal{A}0}\mathcal{V}}{|\gamma_{\mathcal{A}0}|} - \frac{M_{\mathcal{B}0}\mathcal{V}}{|\gamma_{\mathcal{B}0}|} \right] + \hbar \sum_{\pmb{\kappa}} T_{2\pmb{\kappa}} \nonumber \\
  &  + \hbar \sum_{\pmb{\kappa}} T_{3\pmb{\kappa}} ,
\end{align}
where $T_{2\pmb{\kappa}} =  |w_1|^2 - |v_1|^2 + |u_2|^2 - |x_2|^2 + |u_3|^2 - |x_3|^2 + |w_4|^2 - |v_4|^2 $ and $T_{3\pmb{\kappa}} = \{ (1 + 2 ( |w_1|^2 - |x_1|^2 ) \} n_{\alpha_{\pmb{\kappa}}}  - \{ (1 + 2 ( |x_2|^2 - |w_2|^2 ) \} n_{\beta_{\pmb{-\kappa}}} + \{ (1 + 2 ( |u_3|^2 - |v_3|^2 ) \} n_{\alpha_{-\pmb{\kappa}}} - \{ (1 + 2 ( |v_4|^2 - |u_4|^2 ) \} n_{\beta_{\pmb{\kappa}}}$. In Eq. (\ref{eq:spin}) above, the first term denotes the total spin of a fully saturated Neel-ordered state, while the second expresses vacuum fluctuations. The third term represents the contribution of the quasiparticles with $n_{X}$ the number of $\tilde{X}$ quasiparticles, and the curly bracketed prefactor of $n_{X}$ denoting the corresponding quasiparticle spin (in units of $\hbar$). The coefficients associated with squeezing between the same sublattice modes lead to an enhancement in the quasiparticle spin (magnitude), while the ones that effect hybridization between the different sublattice modes cause a reduction in the spin.

We consider three cases - quasiferromagnet, ferrimagnet, and antiferromagnet - that exhibit, disregarding DI,  0, 1 and infinite crossing points on the dispersion diagram. From Eqs. (\ref{eq:sol1}) and (\ref{eq:spin}), we note that the quasiparticle spin in all these cases remains $\pm \hbar$ if DI is disregarded. Together, these three cases encompass all topologically distinct dispersions allowed in our system. Since our model is applicable over a broad domain, we refrain from assigning parameters values corresponding to a particular material, and simply affirm to typical orders of magnitude for the different parameters~\cite{Gomonay2014}. Deferring the specification of the exact values to supplemental material~\cite{SupplMat}, we choose intrasublattice exchange, intersublattice exchange, easy-axis anisotropy, and sublattice saturation magnetization around 1000 T, 100 T, 0.1 T, and 100 kA$/$m~\cite{Gomonay2014}, respectively. We further assume the magnetic specimen to be a film in the y-z plane. 

By quasiferromagnet, we refer to a system with one sublattice dominating over the other [Fig. \ref{fig:ferro}(a)]. As a result, one dispersion branch is much lower in energy than the other and behaves like in a ferromagnet. A prototypical example of such a material is yttrium iron garnet~\cite{Weiler2013,Barker2016,Ritzmann2017}. Figure \ref{fig:ferro}(a) shows that the quasiparticles in a part of this low energy band have spin greater than $\hbar$. This is due to the DI-mediated squeezing of magnons as predicted in ferromagnets~\cite{Kamra2016A}. The high-energy band does not change considerably due to DI since the latter is much weaker than the antiferromagnetic exchange, which dominates this band. 

When both sublattices have comparable properties, for example in gadolinium iron garnet~\cite{Gepraegs2016}, the two bands may (anti)cross as shown in Fig.\ref{fig:ferro}(b). Two primary effects of DI can be seen in the dispersion: \textrm{(i)} the crossing is converted into an anticrossing, and (\textrm{ii}) the corresponding wavevector is shifted owing to the DI contribution to the band energies. Around this anticrossing point [$k \approx \sqrt{\mathcal{J}(|\gamma_{\mathcal{A}}|M_{\mathcal{A}0} - |\gamma_{\mathcal{B}}|M_{\mathcal{B}0})/2(\mathcal{J}_{\mathcal{A}}|\gamma_{\mathcal{A}}|M_{\mathcal{A}0} - \mathcal{J}_{\mathcal{B}} |\gamma_{\mathcal{B}}|M_{\mathcal{B}0})}$], the two kinds of magnons hybridize creating new quasiparticles with spin varying continuously between $\pm \hbar$. We further note that the quasiparticle spin in the low wavenumber region, not depicted in Fig. \ref{fig:ferro}(b), is greater than $\hbar$ due to squeezing.

Antiferromagnets have sublattices with identical properties resulting in, disregarding DI, two degenerate bands. This degeneracy is spontaneously lifted, with the band splitting $\delta f$ in the GHz range, by DI [Fig. \ref{fig:ferro}(c)]. The resulting quasiparticles are superpositions of spin $\pm \hbar$ magnons, and possess zero spin~\footnote{No net spin can also be understood in terms of the symmetry between the two sublattices which enforces $|u_1| = |x_1|, |v_1| = |w_1|$ and so on. Employing such relations in the normalization condition Eq. (\ref{eq:S3}) directly leads to zero spin via Eq. (\ref{eq:spin})}. The band degeneracy can also be lifted by a magnetic field applied along the anisotropy axis which breaks the symmetry between the two sublattices. When this field is much larger than the sublattice saturation magnetizations,  the effects of DI can be disregarded as shown in Fig. \ref{fig:split}.


{\it Discussion.}  While the quasiparticles discussed above have several exotic properties, the emphasis here has been on their classification ($<$ or $>$ $\hbar$) based on average spin. This is partly because their spin is amenable to experimental observation and determines the nature of their interaction with other excitations. Spin transport via these excitations and their experimental detection have been discussed elsewhere~\cite{Kamra2017}.

Equation (\ref{eq:spin}) indicates that non-integral spin and vacuum fluctuations are two facets of the same phenomenon. In particular, squeezing~\cite{Walls2008,Kamra2016A} simultaneously leads to quasiparticle spin larger than $\hbar$ and vacuum fluctuations. We emphasize again that squeezed-excitations can qualitatively be understood within a single sublattice model as coherent superposition of different number states of spin $\hbar$ magnons. Here we discuss suggestive similarities between squeezed ground state and complex frustrated spin-ice systems~\cite{Bramwell2001}. Much like in the latter~\cite{Balents2010}, the essential ingredient to squeezing of the ground state is DI, and not antiferromagnetic exchange. Physically, DI allows pairs of magnons to be spontaneously created or destroyed in the system [see Eq. (\ref{eq:Hamil1})] leading to a degeneracy between different magnon number states. The large degeneracy of the ground state is also considered a defining feature of frustration~\cite{Balents2010}. While squeezing merely reduces the net spin in the ground state, geometrical frustration completely annihilates it. However, squeezing is not dependent upon the system geometry and hence constitutes a different kind of frustration.

\begin{figure}[t]
\begin{center}
\includegraphics[width=85mm]{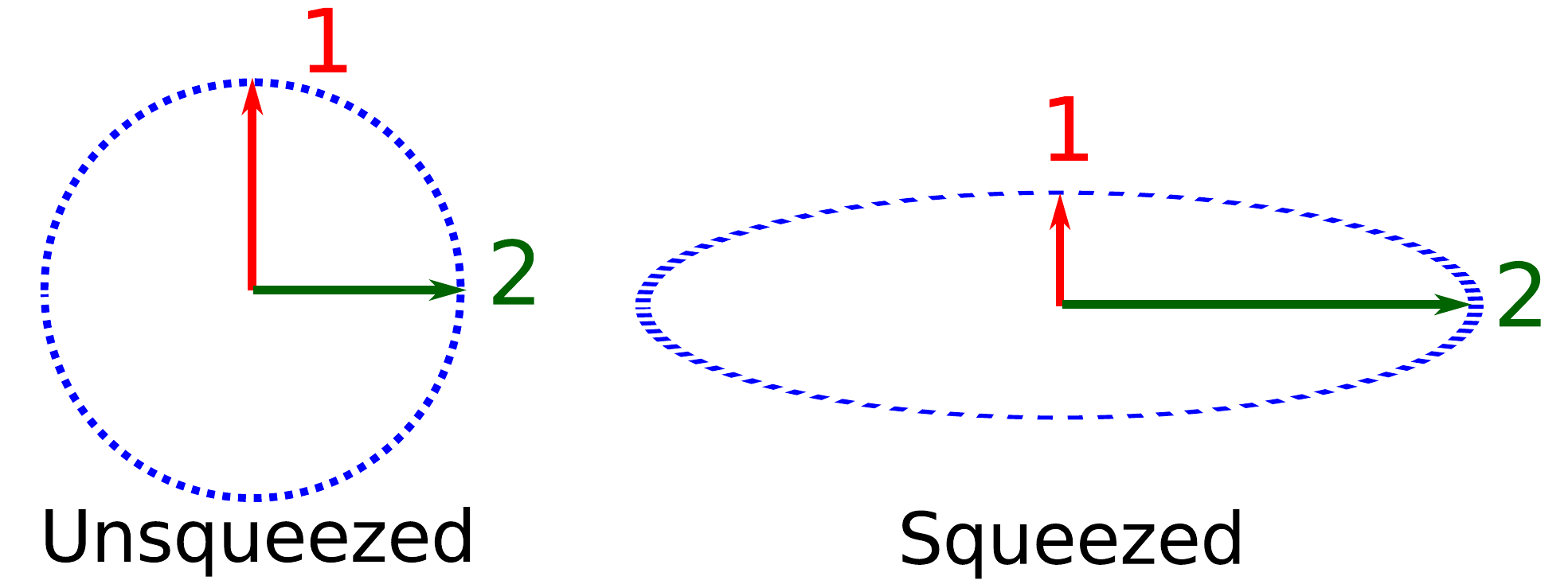}
\caption{Uniform mode squeezing and frustration. In the squeezed state, the total spin vector needs to change its length by adapting the number of excited magnon modes as it goes through different parts of its classical trajectory.}
\label{fig:squeezing}
\end{center}
\end{figure}

The analogy between frustration and squeezing is represented in full generality by the two mathematical conditions discussed above: reduction in the net spin and the large ground state degeneracy. However, we provide a physically intuitive picture for the special case of uniform mode ($\pmb{k} = \pmb{0}$) squeezing. Figure \ref{fig:squeezing} depicts a semi-classical schematic of the net spin vector looking from above. Because of the Heisenberg uncertainty principle, the spin fluctuates about the z-direction which is represented as precessing on the dotted line within a semi-classical picture. Because of a higher energy price in position 1 and the concomitant squeezing, the net spin vector tries to reduce its length by generating magnons spontaneously. Thus, magnons are being generated and destroyed as the spin vector traverses its classical trajectory. In the quantum picture, the spin does not have a trajectory but rather a distribution and the different magnon number states form a coherent superposition. This is reminiscent of the phase dependent noise in squeezing~\cite{Gerry2004}. While geometrical frustration results from conflicting demands on a spin from its geometrical neighbors, squeezing arises due to conflicting demands on the magnon number from different parts of the phase space.


{\it Summary.} We predict exotic bosonic quasiparticles with non-integral average spin in magnets with simple uniformly magnetized ground states. The excitations with spin less and greater than $\hbar$ are, respectively, hybridization and squeezing dominated, and manifest qualitatively different properties. The formation of these quasiparticles requires relaxation of spin conservation in the system, which in our case is provided by the magnetic dipolar interaction. In particular, we find that easy-axis antiferromagnets can host spin-zero quasiparticles. The qualitatively new insights gained into the nature of quasiparticles in antiferromagets fills an important gap in our understanding of these systems, which have recently found their niche in applications~\cite{Jungwirth2016,Wadley2016}. We also suggest a connection between spontaneous squeezing and ``non-geometrical'' frustration which motivates application of quantum optics knowledge in understanding complex solid state systems.

{\it Acknowledgments.} We acknowledge financial support from the Alexander von Humboldt Foundation and the DFG through SFB 767. UA thanks the Belzig group members for their kind hospitality during his visit to Konstanz. 

\bibliography{nonint_arxiv2}

\begin{thebibliography}{44}
\bibitem [{dip()}]{dipnote}%
  \BibitemOpen
  \href@noop {} {}\bibinfo {note} {Here we use the term ``dipolar interaction''
  to represent any contribution to the magnetic Hamiltonian that results in
  spin non-conserving terms up to the second order in the ladder operators.
  Depending upon the material, these terms may predominantly have different
  physical origin such as magnetocrystalline anisotropy, Dzyaloshinksii-Moriya
  interaction and so on.}\BibitemShut {Stop}%
\end{thebibliography}


\widetext
\clearpage
\setcounter{equation}{0}
\setcounter{figure}{0}
\setcounter{table}{0}
\makeatletter
\renewcommand{\theequation}{S\arabic{equation}}

\begin{center}
\textbf{\large Supplementary material with the manuscript Noninteger-spin magnonic excitations in untextured magnets by} \\
\vspace{0.3cm}
Akashdeep Kamra, Utkarsh Agrawal and Wolfgang Belzig
\vspace{0.2cm}
\end{center}

\setcounter{page}{1}

\section{Derivation of the Hamiltonian}

The classical Hamiltonian for the system is given by the integral of energy density over the entire volume $\mathcal{V}$:
\begin{align}
\mathcal{H} & = \int_{\mathcal{V}} d^3 r \ \left(  H_{\mathrm{Z}} + H_{\mathrm{an}} + H_{\mathrm{ex}} + H_{\mathrm{dip}} \right), \\
   & =  \mathcal{H}_{\mathrm{Z}}  + \mathcal{H}_{\mathrm{an}} + \mathcal{H}_{\mathrm{ex}} + \mathcal{H}_{\mathrm{dip}},
\end{align}
with contributions from Zeeman, anisotropy, exchange and dipolar interaction energies. With an applied magnetic field $H_{0}~\hat{\pmb{z}}$, the Zeeman contribution reads:
\begin{align}\label{eq:Z1}
\mathcal{H}_{\mathrm{Z}} & = - \int_{\mathcal{V}} d^3 r \  \mu_0 H_0 (M_{\mathcal{A}z} + M_{\mathcal{B}z}),
\end{align}
where $\pmb{M}_{\mathcal{A}}, \pmb{M}_{\mathcal{B}}$ are the magnetization fields corresponding to the two sublattices. Quantization of Hamiltonian is achieved by replacing the classical variables $\pmb{M}_{\mathcal{A}}, \pmb{M}_{\mathcal{B}}$ with the corresponding quantum operators $\tilde{\pmb{M}}_{\mathcal{A}}, \tilde{\pmb{M}}_{\mathcal{B}}$. The Holstein-Primakoff (HP) transformation~\cite{Holstein1940,Kittel1963} given by:
\begin{align}
\tilde{M}_{\mathcal{A}+}(\pmb{r}) & = \sqrt{2 |\gamma_{\mathcal{A}}| \hbar M_{\mathcal{A}0}} ~ \tilde{a}(\pmb{r}), \label{eq:hp1} \\ 
\tilde{M}_{\mathcal{B}+}(\pmb{r}) & = \sqrt{2 |\gamma_{\mathcal{B}}| \hbar M_{\mathcal{B}0}} ~ \tilde{b}^{\dagger}(\pmb{r}), \\ 
\tilde{M}_{\mathcal{A}\mathrm{z}}(\pmb{r}) & = M_{\mathcal{A}0} - \hbar |\gamma_{\mathcal{A}}| \tilde{a}^{\dagger}(\pmb{r}) \tilde{a}(\pmb{r}), \\
\tilde{M}_{\mathcal{B}\mathrm{z}}(\pmb{r}) & = - M_{\mathcal{B}0} + \hbar |\gamma_{\mathcal{B}}| \tilde{b}^{\dagger}(\pmb{r}) \tilde{b}(\pmb{r}), \label{eq:hp4}
\end{align}
expresses the magnetization in terms of the magnonic ladder operators $\tilde{a}(\pmb{r}),~\tilde{b}(\pmb{r})$ corresponding, respectively, to the two sublattices $\mathcal{A},~\mathcal{B}$. In the above transformation, $\tilde{M}_{P+} = \tilde{M}_{P-}^\dagger = \tilde{M}_{Px} + (\gamma_{P}/|\gamma_{P}|) i \tilde{M}_{Py}$, and $\gamma_{P}$, $M_{\mathcal{P}0}$ are the gyromagnetic ratio and the saturation magnetization corresponding to sublattice P. Employing HP transformation [Eqs. (\ref{eq:hp1}) - (\ref{eq:hp4})] into Eq. (\ref{eq:Z1}), we obtain the Zeeman contribution to the Hamilton operator:
\begin{align}\label{eq:Z}
\tilde{\mathcal{H}}_{\mathrm{Z}} & =\sum_{\pmb{k}} \mu_0 H_0 \hbar (|\gamma_{\mathcal{A}}| \tilde{a}^\dagger_{\pmb{k}} \tilde{a}_{\pmb{k}} - |\gamma_{\mathcal{B}}| \tilde{b}^\dagger_{\pmb{k}} \tilde{b}_{\pmb{k}} ), 
\end{align}
where $\tilde{a}_{\pmb{k}},~\tilde{b}_{\pmb{k}}$ are the magnon operators in the k-space, and we have dropped the constant contribution to the Hamiltonian. 

Proceeding along analogous lines, we now consider the easy-axis anisotropy energy:
\begin{align}
\mathcal{H}_{\mathrm{an}} & =  \int_{\mathcal{V}} d^3 r \left( - K_{\mathrm{u}\mathcal{A}} M_{\mathcal{A}\mathrm{z}}^2 - K_{\mathrm{u}\mathcal{B}} M_{\mathcal{B}\mathrm{z}}^2 \right),
\end{align}
where $K_{\mathrm{u}\mathcal{A}},~K_{\mathrm{u}\mathcal{B}}$ parametrize the energy density~\cite{Akhiezer1968}. On quantization and HP transformation, we obtain the corresponding contribution to the Hamiltonian as:
\begin{align}\label{eq:an}
\tilde{\mathcal{H}}_{\mathrm{an}} & = \hbar \sum_{\pmb{k}}  2 K_{\mathrm{u}\mathcal{A}} |\gamma_{\mathcal{A}}| M_{\mathcal{A}0} \tilde{a}^\dagger_{\pmb{k}} \tilde{a}_{\pmb{k}} + 2 K_{\mathrm{u}\mathcal{B}} |\gamma_{\mathcal{B}}| M_{\mathcal{B}0} \tilde{b}^\dagger_{\pmb{k}} \tilde{b}_{\pmb{k}}.
\end{align} 
The analogous classical and quantum Hamiltonians corresponding to the exchange energy are~\cite{Akhiezer1968}:
\begin{align}
\mathcal{H}_{\mathrm{ex}}  = &  \int_{\mathcal{V}} d^3 r \left[ \sum_{x_i = x,y,z} \left\{ \mathcal{J}_{\mathcal{A}}  \frac{\partial \pmb{M}_{\mathcal{A}}}{\partial x_{i}} \cdot \frac{\partial \pmb{M}_{\mathcal{A}}}{\partial x_{i}} + \mathcal{J}_{\mathcal{B}} \frac{\partial \pmb{M}_{\mathcal{B}}}{\partial x_{i}} \cdot \frac{\partial \pmb{M}_{\mathcal{B}}}{\partial x_{i}}  \right. \right. \nonumber \\ 
      & + \left.  \left. \mathcal{J}_{\mathcal{AB}}  \frac{\partial \pmb{M}_{\mathcal{A}}}{\partial x_{i}} \cdot \frac{\partial \pmb{M}_{\mathcal{B}}}{\partial x_{i}} \right\} + \mathcal{J} \pmb{M}_{\mathcal{A}} \cdot \pmb{M}_{\mathcal{B}} \right], \\
\tilde{\mathcal{H}}_{\mathrm{ex}}  = & \hbar \sum_{\pmb{k}} \left[ 2 \mathcal{J}_{\mathcal{A}} k^2 |\gamma_{\mathcal{A}}| M_{\mathcal{A}0} + \mathcal{J} |\gamma_{\mathcal{B}}| M_{\mathcal{B}0} \right] \tilde{a}^\dagger_{\pmb{k}} \tilde{a}_{\pmb{k}}  \nonumber \\
  &  + \hbar \sum_{\pmb{k}} \left[ 2 \mathcal{J}_{\mathcal{B}} k^2 |\gamma_{\mathcal{B}}| M_{\mathcal{B}0} + \mathcal{J} |\gamma_{\mathcal{A}}| M_{\mathcal{A}0} \right] \tilde{b}^\dagger_{\pmb{k}} \tilde{b}_{\pmb{k}} \nonumber \\
    & + \hbar \sum_{\pmb{k}} \left[  \mathcal{J}_{\mathcal{AB}} k^2 \sqrt{|\gamma_{\mathcal{A}}| M_{\mathcal{A}0} |\gamma_{\mathcal{B}}| M_{\mathcal{B}0}} + \mathcal{J} \sqrt{|\gamma_{\mathcal{A}}| M_{\mathcal{A}0} |\gamma_{\mathcal{B}}| M_{\mathcal{B}0}} \right] \left( \tilde{a}_{\pmb{k}} \tilde{b}_{-\pmb{k}} + \tilde{a}_{\pmb{k}}^\dagger \tilde{b}_{-\pmb{k}}^\dagger \right), \label{eq:ex}
\end{align}
where $\mathcal{J}_{\mathcal{A}}$, $\mathcal{J}_{\mathcal{B}}$ parametrize the intrasublattice exchange while $\mathcal{J}_{\mathcal{AB}}$, $\mathcal{J}$ parametrize intersublattice exchange interaction.

The dipolar interaction is treated within a mean field approximation via the so-called demagnetization field $\pmb{H}_m$ generated by the magnetization:
\begin{align}
 \mathcal{H}_{\mathrm{dip}} & = -  \int_{\mathcal{V}} d^3 r \ \frac{1}{2} \mu_0 \pmb{H}_m \cdot \pmb{M},
\end{align}
with $\pmb{M} = \pmb{M}_{\mathcal{A}0} + \pmb{M}_{\mathcal{B}0}$ the total magnetization. The magnetization and the demagnetization field are split into spatially uniform and non-uniform contributions $\pmb{H}_m   = \pmb{H}_{u} + \pmb{H}_{nu}$ and $\pmb{M}   = \pmb{M}_{u} + \pmb{M}_{nu}$ thereby affording the following relation between the uniform components~\cite{Akhiezer1968,Kittel1963}:
\begin{align}
\pmb{H}_{u} = - N_x M_{ux}~ \hat{\pmb{x}} -  N_y M_{uy} ~\hat{\pmb{y}} - N_z M_{uz} ~\hat{\pmb{z}},
\end{align} 
where $N_{x,y,z}$ are the eigenvalues of the demagnetization tensor which is diagonal in the chosen coordinate system. Within the magnetostatic approximation, the non-uniform components obey the equations~\cite{Akhiezer1968,Kittel1963}:
\begin{align}
\pmb{\nabla} \times \pmb{H}_{nu} & = 0, \\
\pmb{\nabla} \cdot \left( \pmb{H}_{nu} + \pmb{M}_{nu} \right) & = 0. \label{divdemagfield}
\end{align}
Employing the equations above and HP transformation [Eqs. (\ref{eq:hp1}) - (\ref{eq:hp4})], straightforward albeit lengthy algebra leads to the following expression for the dipolar contribution to the Hamiltonian:
\begin{align}\label{eq:dip}
\tilde{\mathcal{H}}_{\mathrm{dip}}  = & \hbar \mu_0 |\gamma_{\mathcal{A}}| \sum_{\pmb{k}}  \left[ N_z (M_{\mathcal{B}0} - M_{\mathcal{A}0})  + \delta_{\pmb{k},\pmb{0}} \frac{N_x + N_y}{2} M_{\mathcal{A}0} + (1 - \delta_{\pmb{k},\pmb{0}}) \frac{\sin^2\left(\theta_{\pmb{k}}\right)}{2} M_{\mathcal{A}0} \right] \tilde{a}^\dagger_{\pmb{k}} \tilde{a}_{\pmb{k}} \nonumber \\
   & + \hbar \mu_0 |\gamma_{\mathcal{B}}| \sum_{\pmb{k}}  \left[ N_z (M_{\mathcal{A}0} - M_{\mathcal{B}0})  + \delta_{\pmb{k},\pmb{0}} \frac{N_x + N_y}{2} M_{\mathcal{B}0} + (1 - \delta_{\pmb{k},\pmb{0}}) \frac{\sin^2\left(\theta_{\pmb{k}}\right)}{2} M_{\mathcal{B}0} \right] \tilde{b}^\dagger_{\pmb{k}} \tilde{b}_{\pmb{k}} \nonumber \\
   & + \hbar \mu_0 \sqrt{|\gamma_{\mathcal{A}}| M_{\mathcal{A}0} |\gamma_{\mathcal{B}}| M_{\mathcal{B}0}} \sum_{\pmb{k}}  \left[ \delta_{\pmb{k},\pmb{0}} \frac{N_x + N_y}{2} + (1 - \delta_{\pmb{k},\pmb{0}}) \frac{\sin^2\left(\theta_{\pmb{k}}\right)}{2} \right] \left( \tilde{a}_{\pmb{k}} \tilde{b}_{-\pmb{k}} + \tilde{a}^\dagger_{\pmb{k}} \tilde{b}^{\dagger}_{-\pmb{k}} \right) \nonumber \\
   & + \left\{ \hbar \mu_0 |\gamma_{\mathcal{A}}| M_{\mathcal{A}0}  \sum_{\pmb{k}}  \left[ \delta_{\pmb{k},\pmb{0}} \frac{N_x - N_y}{4} + (1 - \delta_{\pmb{k},\pmb{0}}) \frac{\sin^2\left(\theta_{\pmb{k}}\right)}{4} e^{i 2 \phi_{\pmb{k}}} \right]  \tilde{a}_{\pmb{k}} \tilde{a}_{-\pmb{k}} \right\}  + \mathrm{h.c.} \nonumber \\
   & + \left\{ \hbar \mu_0 |\gamma_{\mathcal{B}}| M_{\mathcal{B}0}  \sum_{\pmb{k}}  \left[ \delta_{\pmb{k},\pmb{0}} \frac{N_x - N_y}{4} + (1 - \delta_{\pmb{k},\pmb{0}}) \frac{\sin^2\left(\theta_{\pmb{k}}\right)}{4} e^{- i 2 \phi_{\pmb{k}}} \right]  \tilde{b}_{\pmb{k}} \tilde{b}_{-\pmb{k}} \right\}  + \mathrm{h.c.} \nonumber \\
   & + \left\{ \hbar \mu_0 \sqrt{|\gamma_{\mathcal{A}}| M_{\mathcal{A}0} |\gamma_{\mathcal{B}}| M_{\mathcal{B}0}}   \sum_{\pmb{k}}  \left[ \delta_{\pmb{k},\pmb{0}} \frac{N_x - N_y}{2} + (1 - \delta_{\pmb{k},\pmb{0}}) \frac{\sin^2\left(\theta_{\pmb{k}}\right)}{2} e^{i 2 \phi_{\pmb{k}}} \right]  \tilde{a}_{\pmb{k}} \tilde{b}_{\pmb{k}}^\dagger \right\}  + \mathrm{h.c.} ~~,
\end{align} 
where $\theta_{\pmb{k}}$ and $\phi_{\pmb{k}}$ are the polar and azimuthal angles of $\pmb{k}$.

Combining Eqs. (\ref{eq:Z}), (\ref{eq:an}), (\ref{eq:ex}) and (\ref{eq:dip}), the full Hamiltonian reads:
\begin{align}
\tilde{\mathcal{H}}  = & \sum_{\pmb{k}} \left[ \frac{A_{\pmb{k}}}{2} \tilde{a}_{\pmb{k}}^{\dagger} \tilde{a}_{\pmb{k}} + \frac{B_{\pmb{k}}}{2} \tilde{b}_{\pmb{k}}^{\dagger} \tilde{b}_{\pmb{k}} + C_{\pmb{k}} \tilde{a}_{\pmb{k}} \tilde{b}_{-\pmb{k}} + D_{\pmb{k}} \tilde{a}_{\pmb{k}} \tilde{a}_{-\pmb{k}} + E_{\pmb{k}} \tilde{b}_{\pmb{k}} \tilde{b}_{-\pmb{k}} + F_{\pmb{k}} \tilde{a}_{\pmb{k}} \tilde{b}_{\pmb{k}}^{\dagger} \right] + \mathrm{h.c.} \quad ,
\end{align}
where
\begin{align}
\frac{A_{\pmb{k}}}{\hbar} = & \mu_0 H_0 |\gamma_{\mathcal{A}}| +  2 K_{\mathrm{u}\mathcal{A}} |\gamma_{\mathcal{A}}| M_{\mathcal{A}0} + 2 \mathcal{J}_{\mathcal{A}} k^2 |\gamma_{\mathcal{A}}| M_{\mathcal{A}0} + \mathcal{J} |\gamma_{\mathcal{B}}| M_{\mathcal{B}0} \nonumber \\
    & +  \mu_0 |\gamma_{\mathcal{A}}|  \left[ N_z (M_{\mathcal{B}0} - M_{\mathcal{A}0})  + \delta_{\pmb{k},\pmb{0}} \frac{N_x + N_y}{2} M_{\mathcal{A}0} + (1 - \delta_{\pmb{k},\pmb{0}}) \frac{\sin^2\left(\theta_{\pmb{k}}\right)}{2} M_{\mathcal{A}0} \right] , \\
\frac{B_{\pmb{k}}}{\hbar} = & - \mu_0 H_0 |\gamma_{\mathcal{B}}| +  2 K_{\mathrm{u}\mathcal{B}} |\gamma_{\mathcal{B}}| M_{\mathcal{B}0} + 2 \mathcal{J}_{\mathcal{B}} k^2 |\gamma_{\mathcal{B}}| M_{\mathcal{B}0} + \mathcal{J} |\gamma_{\mathcal{A}}| M_{\mathcal{A}0} \nonumber \\
    & +  \mu_0 |\gamma_{\mathcal{B}}|  \left[ N_z (M_{\mathcal{A}0} - M_{\mathcal{B}0})  + \delta_{\pmb{k},\pmb{0}} \frac{N_x + N_y}{2} M_{\mathcal{B}0} + (1 - \delta_{\pmb{k},\pmb{0}}) \frac{\sin^2\left(\theta_{\pmb{k}}\right)}{2} M_{\mathcal{B}0} \right], \\
 \frac{C_{\pmb{k}}}{\hbar} = & \sqrt{|\gamma_{\mathcal{A}}| M_{\mathcal{A}0} |\gamma_{\mathcal{B}}| M_{\mathcal{B}0}} \left[ \mathcal{J} + \mathcal{J}_{\mathcal{AB}} k^2 +  \mu_0 \delta_{\pmb{k},\pmb{0}} \frac{N_x + N_y}{2} + \mu_0 (1 - \delta_{\pmb{k},\pmb{0}}) \frac{\sin^2\left(\theta_{\pmb{k}}\right)}{2}   \right], \\
 \frac{D_{\pmb{k}}}{\hbar} = & \mu_0 |\gamma_{\mathcal{A}}| M_{\mathcal{A}0} \left[ \delta_{\pmb{k},\pmb{0}} \frac{N_x - N_y}{4} + (1 - \delta_{\pmb{k},\pmb{0}}) \frac{\sin^2\left(\theta_{\pmb{k}}\right)}{4} e^{i 2 \phi_{\pmb{k}}} \right], \\
  \frac{E_{\pmb{k}}}{\hbar} = & \mu_0 |\gamma_{\mathcal{B}}| M_{\mathcal{B}0} \left[ \delta_{\pmb{k},\pmb{0}} \frac{N_x - N_y}{4} + (1 - \delta_{\pmb{k},\pmb{0}}) \frac{\sin^2\left(\theta_{\pmb{k}}\right)}{4} e^{- i 2 \phi_{\pmb{k}}} \right], \\
  F_{\pmb{k}} = & 2 \sqrt{D_{\pmb{k}} E_{\pmb{k}}^*} .
\end{align}

\section{Values of model parameters}
\begin{center}
\begin{tabular}{c | c | c | c | c}
\hline
Parameter & Quasiferromagnet & Ferrimagnet & Antiferromagnet & Units \\
\hline
$\mu_0 H_0$ & 0.05   & 0.05  & 0 & T \\
$N_x$, $N_y$, $N_z$ & 1,0,0  & 1,0,0 & 1,0,0 & Dimensionless \\
$\gamma_{\mathcal{A}}$  & 1.8 & 1.8  & 1.8 & $\times 10^{11}$ $\mathrm{s}^{-1} \mathrm{T}^{-1}$  \\
$\gamma_{\mathcal{B}}$ & 1.8 & 1.8  & 1.8 & $\times 10^{11}$ $\mathrm{s}^{-1} \mathrm{T}^{-1}$ \\
$M_{\mathcal{A}}$ & 5  & 5 & 5 & $\times 10^5$ A/m \\
$M_{\mathcal{B}}$  & 1  & 2.5 & 5 & $\times 10^5$ A/m \\
$\mathcal{J}_{\mathcal{A}}$  &  1 & 5 &  1 & $\times 10^{-23}$ $\mathrm{J}\cdot\mathrm{m} \mathrm{A}^{-2} $   \\
$\mathcal{J}_{\mathcal{B}}$ &  1 & 1 &  1 & $\times 10^{-23}$ $\mathrm{J}\cdot\mathrm{m} \mathrm{A}^{-2} $  \\
$\mathcal{J}_{\mathcal{AB}}$  &  0.1 & 0.1 &  0.1 & $\times 10^{-23}$ $\mathrm{J}\cdot\mathrm{m} \mathrm{A}^{-2} $  \\
$\mathcal{J}$  & 1  & 1 &  5 & $\times 10^{-4}$ $\mathrm{J} \mathrm{m}^{-1} \mathrm{A}^{-2} $ \\
$K_{u\mathcal{A}}$ & 2 & 2  & 2 & $ \times 10^{-7}$ $\mathrm{J} \mathrm{m}^{-1} \mathrm{A}^{-2} $ \\
$K_{u\mathcal{B}}$  & 2 & 2  & 2 & $ \times 10^{-7}$ $\mathrm{J} \mathrm{m}^{-1} \mathrm{A}^{-2} $ \\
\hline
\end{tabular}
\end{center}

\end{document}